\documentclass[a4,aps,prl,showpacs,amsmath,floatfix,twocolumn]{revtex4}
\usepackage{amssymb}
\usepackage{braket}
\usepackage{graphicx}
\usepackage[colorlinks,linkcolor=blue,anchorcolor=red,citecolor=green]{hyperref}
\usepackage{pifont}
\usepackage{fontawesome} 
\usepackage{comment}
\usepackage{MnSymbol}
\usepackage{array, makecell}

\newcolumntype{L}[1]{>{\raggedright\let\newline\\\arraybackslash\hspace{0pt}}m{#1}}
\newcolumntype{C}[1]{>{\centering\let\newline\\\arraybackslash\hspace{0pt}}m{#1}}
\newcolumntype{R}[1]{>{\raggedleft\let\newline\\\arraybackslash\hspace{0pt}}m{#1}}

\newcommand{\be}{\begin{equation}}
\newcommand{\ee}{\end{equation}}
\newcommand{\bea}{\begin{eqnarray}}
\newcommand{\eea}{\end{eqnarray}}

\newcommand{\ra}{\rangle}

\renewcommand{\vec}[1]{{\bf #1}}

\usepackage[dvipsnames]{xcolor}

\begin{document}

\title{Nonequilibrium Exchange Nonlinear Hall Effect}

\author{John Tan$^{1,\dagger}$, Oles Matsyshyn$^{1,\dagger,*}$, Giovanni Vignale$^{2}$, and Justin C. W. Song$^{1}$}
\email{oles.matsyshyn@ntu.edu.sg; justinsong@ntu.edu.sg}

\affiliation{$^{1}$Division of Physics and Applied Physics, School of Physical and Mathematical Sciences, Nanyang Technological University, Singapore 637371}
\affiliation{$^{2}$ The Institute for Functional Intelligent Materials (I-FIM),
National University of Singapore, 4 Science Drive 2, Singapore 117544}

\begin{abstract}
Quantum geometric electronic responses are often viewed through a non-interacting lens: independent quasiparticles accumulate Berry phases as they move through a static crystal and background potential. Here we argue that the combined action of electron-electron interactions and an out-of-equilibrium many-body state can produce striking departures from this familiar picture. We demonstrate how nonequilibrium exchange interactions produce a nonequilibrium collective quantum geometry distinct from that of its equilibrium ground state. We find this manifests as an exchange induced nonlinear Hall effect with nonlinear Hall current signals competitive with that of well-known non-interacting mechanisms. This highlights the critical role electron interactions and nonequilibrium states can play in the nonlinear response of quantum matter.
\end{abstract}

\maketitle

Momentum-space quantum geometry of Bloch bands play a surprisingly critical role in nonlinear response \cite{Moore2010,Sodemann2015,Gao2014,Ahn2022}. A prime example are the momentum-space Berry curvature dipole \cite{Sodemann2015} and the Berry connection polarizability \cite{Gao2014,Liu2021,Wang2021} of electrons which endow them with a nonlinear Hall response. This geometry is often viewed as a characteristic of individual Bloch electrons at each momentum $\mathbf{k}$. As a result, nonlinear Hall currents are used as a sensitive diagnostic of the momentum-space geometry of Bloch bands in quantum matter \cite{ma2019,Kang2019,Gao2023,Wang2023}.

\begin{figure}[t]
    \centering
    \includegraphics[width=0.5\textwidth]{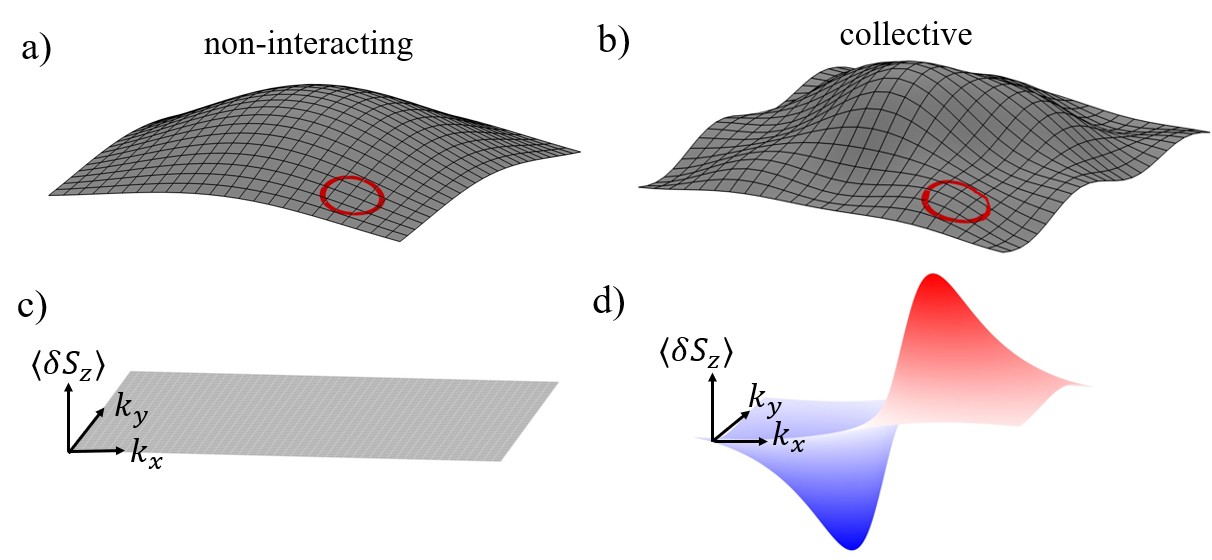}
    \caption{\textbf{Nonequilibrium collective quantum geometry} (a) Bloch wavefunction variation with $\mathbf{k}$ characterizes its momentum-space quantum geometry; e.g., Berry phase across small (red) loops track momentum-space Berry curvature. (c) In non-interacting materials, these are locked to the equilibrium state, e.g., equilibrium spin density in a PT symmetric antiferromagnet vanishes. (b) When pushed out-of-equilibrium by an applied electric field, a {\it collective} quantum geometry develops, mirroring the nonequilibrium state e.g., tracked by (d) a nonequilibrium spin density.} 
    \label{Fig 1}
\end{figure}

Here we argue that electron interactions enable a nonlinear Hall effect to arise {\it collectively}. In particular, we find that when electrons are pushed out-of-equilibrium by an electric field, {\it nonequilibrium} exchange interactions produce a collective quantum geometry distinct from its non-interacting value (Fig \ref{Fig 1}a,b). Unlike non-interacting momentum-space quantum geometry that is generated by momentum translations (e.g., through Berry connections~\cite{Ahn2022}), the collective quantum geometry induced by nonequilibrium exchange is generated by spin. As we will see below, this manifests as an exchange nonlinear Hall (ENH) effect with magnitude that is tuned by the strength of electron interactions and sensitive to the collective spin-density profile. 

Importantly, ENH includes interactions beyond a ``frozen'' equilibrium mean-field. This distinction is particularly striking in Parity-Time (PT) symmetric magnets where PT symmetry ensures spin density at every momentum $\mathbf{k}$ vanishes (Fig \ref{Fig 1}c) \cite{Ma2023} thereby suppressing equilibrium exchange energy. In contrast, when it is pushed out-of-equilibrium by an electric field, spin density accumulates (Fig \ref{Fig 1}d) generating a nonequilibrium exchange interaction~\cite{Vignale2008}.

At a fundamental level, our work demonstrates a sharp distinction between linear and nonlinear responses. Responses such as 
the linear Hall effect can be readily described purely in terms of the momentum-space quantum geometry of an equilibrium interacting bandstructure~\cite{Xiao2010,ThoulessNiu1985}. In contrast, we find that the nonlinear Hall effect cannot: ENH depends on an interacting nonequilibrium quantum geometry arising from exchange interactions absent in equilibrium. Our work sits in a recent surge of interest in the novel feedback between a (driven) nonequilibrium many-body state \cite{Rudner2019,Mazza2019,Balram2019,Guerci2020,Claassen2023,Ying2024,FFL,UCFFL} and interactions that produce a range of unconventional driven phases \cite{Rudner2019,Mazza2019,Guerci2020,Ying2024, FFL, UCFFL} and quantum transport behavior \cite{Balram2019, FFL, UCFFL}. For instance, nonequilibrium exchange interactions can enable current-induced gaps in topological insulators~\cite{Balram2019}; population inversion, combined with electron interactions, can produce ``antiscreening'' and ferroelectric like states \cite{Ying2024}. For the second-order nonlinear Hall effect we focus on here, we find that nonequilibrium ENH is sizeable with signals that can rival their more familiar non-interacting counterparts\cite{Sodemann2015,Gao2014,Liu2021,Wang2021}. This renders ENH a critical and distinct exemplar of a growing many-body quantum geometry.

\vspace{2mm}

\textit{\color{blue} Nonequilibrium exchange interaction.} We begin by examining the impact of electron interactions within an effective mean-field description: $\sum_{n,m,\mathbf{k}} c_{n,\mathbf{k}} ^\dagger\hat{H}_{\rm MF} \,  c_{m,\mathbf{k}} =  \sum_{\mathbf{k}, n,m} c^\dagger_{n,\mathbf{k}} \hat h (\mathbf{k}) c_{m,\mathbf{k}} + H_{\rm e-e}$ where $\hat{h}(\mathbf{k})$ is the crystalline Bloch Hamiltonian and $H_{\rm e-e}$ is a self-energy 
\begin{equation}
H_{\rm e-e} =
 -\hspace{-3mm}\sum_{\{\boldsymbol{n}\},\mathbf{k,k'}}{V}_{\mathbf{k'-k}}\Gamma^{\{\boldsymbol{n}\}}_{\mathbf{k',k}}\rho_{n_3n_1}({\mathbf{k'})}\hat{c}^\dagger_{n_2,\mathbf{k}}\hat{c}_{n_4,\mathbf{k}} + H_{\rm d},\label{HF}
\end{equation}
where $\hat c^\dagger_{n,\mathbf{k}}$ is a Bloch creation operator of the Bloch state $|n, \mathbf{k}\rangle$ with $V_{\mathbf{q}}$ being the Fourier harmonic of the interaction potential, the overlap factor reads $\Gamma^{\{\boldsymbol{n}\}}_{\mathbf{k},\mathbf{k'}} = \sum_{\sigma, \sigma'} \langle n_1, \mathbf{k'}| n_4, \mathbf{k} \rangle_\sigma \langle n_2, \mathbf{k}| n_3, \mathbf{k'} \rangle_{\sigma'}$ where we have explicitly specified its spin-resolution with spin indices $\sigma, \sigma' = \{\uparrow, \downarrow\}$ to emphasize the spin physics we will discuss below; here we grouped the band indices as $\mathbf{n}=(n_1,n_2,n_3,n_4)$ for compactness. The generalized mean-field density is $\rho_{n_3n_1}({\mathbf{k'}})\equiv\langle \hat{c}^\dagger_{n_1,\mathbf{k'}}\hat{c}_{n_3,\mathbf{k}'}\rangle$ and $H_d$ is the direct interaction. Such mean-field densities encode collective behavior and allow to describe a diverse range of interacting phases in a physically transparent way. 

At equilibrium, $\rho_{mn}(\mathbf{k}) =  \rho_{mn}^{(0)} (\mathbf{k})$ is fixed producing a static $\hat{H}_{\rm MF} (\rho_{mn}^{(0)}, \mathbf{k}) = \hat{H}_{\rm MF}^{(0)} (\mathbf{k})$. Such static $\hat{H}_{\rm MF}^{(0)} (\mathbf{k})$ are often used to track the equilibrium bandstructure where $\hat{H}_{\rm MF}^{(0)} (\mathbf{k}) \, |m, \mathbf{k}\rangle = \epsilon_m (\mathbf{k}) \, |m, \mathbf{k}\rangle$ describes quasiparticle eigenstates; these in turn enable to model equilibrium properties such as magnetization \cite{Shi2007} and the quantum geometry of materials \cite{Ahn2022,Xiao2010}. 

However, when an uniform electric field $\mathbf{E}$ is applied to this equilibrium, the mean-field densities $\rho_{mn}(\mathbf{k}) =\rho_{mn}^{(0)}(\mathbf{k}) + \delta \rho_{mn}(\mathbf{k})$ become dynamical \cite{Vignale2008}. To demonstrate this systematically, we first write the electric field perturbation as $H' = e\hat{\mathbf{r}} \cdot \mathbf{E}$ with $\hat{\mathbf{r}}  = i \nabla_{\mathbf{k}}$ the position operator and obtain the density matrix to first order in the applied field~\cite{Parker2019,Oles2021} 
\begin{equation}\label{eq:MF1}
        \delta\rho_{nm} ({\mathbf{k}})=-\frac{e\tau}{\hbar}E^{\alpha}\delta_{nm}\partial^\alpha_\mathbf{k}f_n+e{E}^\alpha r^\alpha_{nm}(\mathbf{k})\frac{f_{mn}}{\epsilon_{nm}},
\end{equation}
where $\tau$ is a relaxation time, $\epsilon_{nm}\equiv \epsilon_n(\mathbf{k})-\epsilon_m(\mathbf{k})$ and $f_{nm}\equiv f[\epsilon_n(\mathbf{k})]-f[\epsilon_m(\mathbf{k})]$ for Fermi-Dirac function $f[x]$; summation over repeated spatial indices $\alpha$ is implied. Here the matrix element of the operator $\hat{\mathcal{O}}$ reads $\mathcal{O}_{nm}(\mathbf{k})=\langle n,\mathbf{k}|\hat{\mathcal{O}}|m,\mathbf{k}\rangle$. The first term of Eq.~(\ref{eq:MF1}) arises from a shift in the Fermi surface and scales as $\tau$ while the second term arises from off-diagonal coherences between bands produced by the applied $\mathbf{E}$; the latter does not scale with $\tau$.

Plugging Eq.~(\ref{eq:MF1}) into Eq.~(\ref{HF}), we can estimate the $\mathbf{E}$ driven change in the exchange energy $\delta H_{\rm e-e} = \sum_{n,m,k} c^\dagger_{n,\mathbf{k}} \delta \hat{U}_{\rm ex} c_{m,\mathbf{k}}$ within a first Born approximation~\cite{Vignale2008, Balram2019} as $\delta \hat{U}_{\rm ex} =  U^\eta_{\mathbf{k}} (\mathbf{E})\hat{s}^\eta$ with  
\begin{equation}
\hspace{-2mm} U^\eta_{\mathbf{k}} (\mathbf{E})= - \hspace{-2mm}\sum_{\mathbf{k'},n,m} \hspace{-1mm}{V}_{\mathbf{k'-k}}s^\eta_{nm}(\mathbf{k'})\delta\rho_{mn}(\mathbf{k'}), \label{eq:selfenergy}
\end{equation}
where $\eta = \{0, 1, 2, 3\}$ with $\hat{s}^{\{1,2,3\}}$ the Pauli spin operators and $\hat{s}^{0} = 1$, the matrix elements $s^\eta_{nm} (\mathbf{k'}) = \langle n, \mathbf{k'} | \hat{s}^\eta| m, \mathbf{k'}\rangle $, and we have used the spin identity $\sum_{\sigma, \sigma'} \langle n_1, \mathbf{k'}| n_4,\mathbf{k} \rangle_\sigma \langle n_2, \mathbf{k}| n_3, \mathbf{k'} \rangle_{\sigma'} =   s^\eta_{n_1n_3}(\mathbf{k'})s^\eta_{n_2n_4}(\mathbf{k})$. In obtaining Eq.~(\ref{eq:selfenergy}) we have noted that an applied uniform electric field does not change the total charge density. As a result, applied uniform $\mathbf{E}$ does not change $H_{\rm d}$ since the Hartree energy depends on total density~\cite{Vignale2008}. 

In PT symmetric materials, such $\mathbf{E}$ driven exchange is particularly dramatic: at equilibrium, PT symmetry ensures that the spin expectation value at every $\mathbf{k}$ is compensated \cite{Ma2023} zeroing the spin contribution to exchange energy. However, when pushed out-of-equilibrium by $\mathbf{E}$, spin density (at each $\mathbf{k}$) can accumulate activating a nonequilibrium exchange interaction. 

\vspace{2mm}
\textit{\color{blue} Nonequilibrium collective quantum geometry.} As we now argue, the nonequilibrium exchange energy induces modifications to the Bloch band quantum geometry. We explicitly examine the $\mathbf{E}$ driven mean-field Bloch Hamiltonian including exchange interaction corrections 
\begin{equation}\label{eq:HMF}
    \hat{H}_{\rm MF} (\mathbf{k})= \hat{H}_{\rm MF}^{(0)}(\mathbf{k}) + e\hat{\mathbf{r}} \cdot \mathbf{E} +  U^\eta_{\mathbf{k}} (\mathbf{E}) \hat{s}^\eta, 
\end{equation}
where $e\hat{\mathbf{r}} \cdot \mathbf{E}$ acts locally in $\mathbf{k}$ while the exchange term is non-local and collective since $ U^\eta_{\mathbf{k}}(\mathbf{E})$ in Eq.~(\ref{eq:selfenergy}) depends on the distribution of $s_{nm}^\eta (\mathbf{k'})$. 

To appreciate how interactions modify the quantum geometry and bandstructure, consider the fundamental building block of Bloch quantum geometry: the intraband Berry connection $\mathbf{\cal A}_n (\mathbf{k})= \langle n, \mathbf{k} | i \nabla_{\mathbf{k}}| n, \mathbf{k}\rangle$. Expanding the Bloch states to first order in $U_{\mathbf{k}}^\eta (\mathbf{E})$ and $\mathbf{E}$ produces a change in the Berry connection as $\delta \mathbf{\cal A}_{n,\mathbf{k}}^\alpha =U^{\eta}_{\mathbf{k}} (\mathbf{E}) \mathcal{S}^{ \eta\alpha }_{n,\mathbf{k}}  + eE^\beta G_{n,{\mathbf{k}}}^{\beta\alpha } $ where 
\begin{equation} 
\mathcal{S}^{ \eta\alpha}_{n,\mathbf{k}}= 2\text{Re}\hspace{-3mm}\sum_{m, \epsilon_n \neq \epsilon_m}\hspace{-3mm} \frac{s_{nm}^{\eta}(\mathbf{k})r_{mn}^\alpha (\mathbf{k})}{\epsilon_{nm}(\mathbf{k})}, 
\label{eq:deltaA}
\end{equation}
with $\mathcal{S}^{ \eta\alpha}_{n,\mathbf{k}}$ is the spin-Berry polarizability (SBP) that captures how the spin distribution alters the Berry connection and $G^{\alpha\beta}_{n,\mathbf{k}} = 2{\rm Re}\left[\sum_m^{\epsilon_n\neq\epsilon_m}r^\alpha_{nm}r^\beta_{mn}/\epsilon_{nm}\right]$ is the familiar Berry connection polarizability (BCP) \cite{Gao2014}. Both $\mathcal{S}^{ \eta \alpha}$ and $G^{\alpha\beta}$ are gauge invariant; notice here that in writing $\epsilon_n\neq\epsilon_m$ we have accounted for degenerate systems ~\cite{Sakurai_Napolitano_2020}.

Note that the non-interacting BCP is expressed in terms of momentum space quantum geometry of its equilibrium bandstructure [e.g., interband Berry connection $r_{nm} (\mathbf{k})$]. It arises even at the non-interacting level where $\mathbf{E}$-field induces mixing between two Bloch states $|n, \mathbf{k}\ra$ and $|m, \mathbf{k}\ra$ at equilibrium. Such coherences~\cite{Gao2014} are critical to the dielectric polarizability of quantum materials~\cite{Komissarov2024}. 

In contrast, the SBP contribution to $\delta A_{n,\mathbf{k}}^\alpha$ is distinct: it is {\it collective} and {\it nonequilibrium}. To see this, notice that the nonequilibrium exchange energy $U_{\vec k}^\eta (\mathbf{E})$ in Eq.~(\ref{eq:selfenergy}) depends on the full spin density distribution of electrons across all $\mathbf{k}$: spin density at $\mathbf{k'} \neq \mathbf{k}$ affect the polarizability at $\mathbf{k}$. This non-locality encodes the interactions of a collective soup of electrons. Similarly, the SBP contribution cannot be expressed simply as a function of the equilibrium self-energy. Instead, it arises from how $U_{\vec k}^\eta (\mathbf{E})$ varies with applied $\mathbf{E}$ and directly depends on the nonequilibrium distribution of electrons in Eq.~(\ref{eq:MF1}). When $U_{\vec k}^\eta (\mathbf{E})$ is combined with the SBP tensor $\mathcal{S}^{\eta \alpha}_{n,\mathbf{k}}$, it produces a nonequilibrium collective quantum geometry.

\begin{figure}[t]
    \centering
    \includegraphics[width=0.48\textwidth]{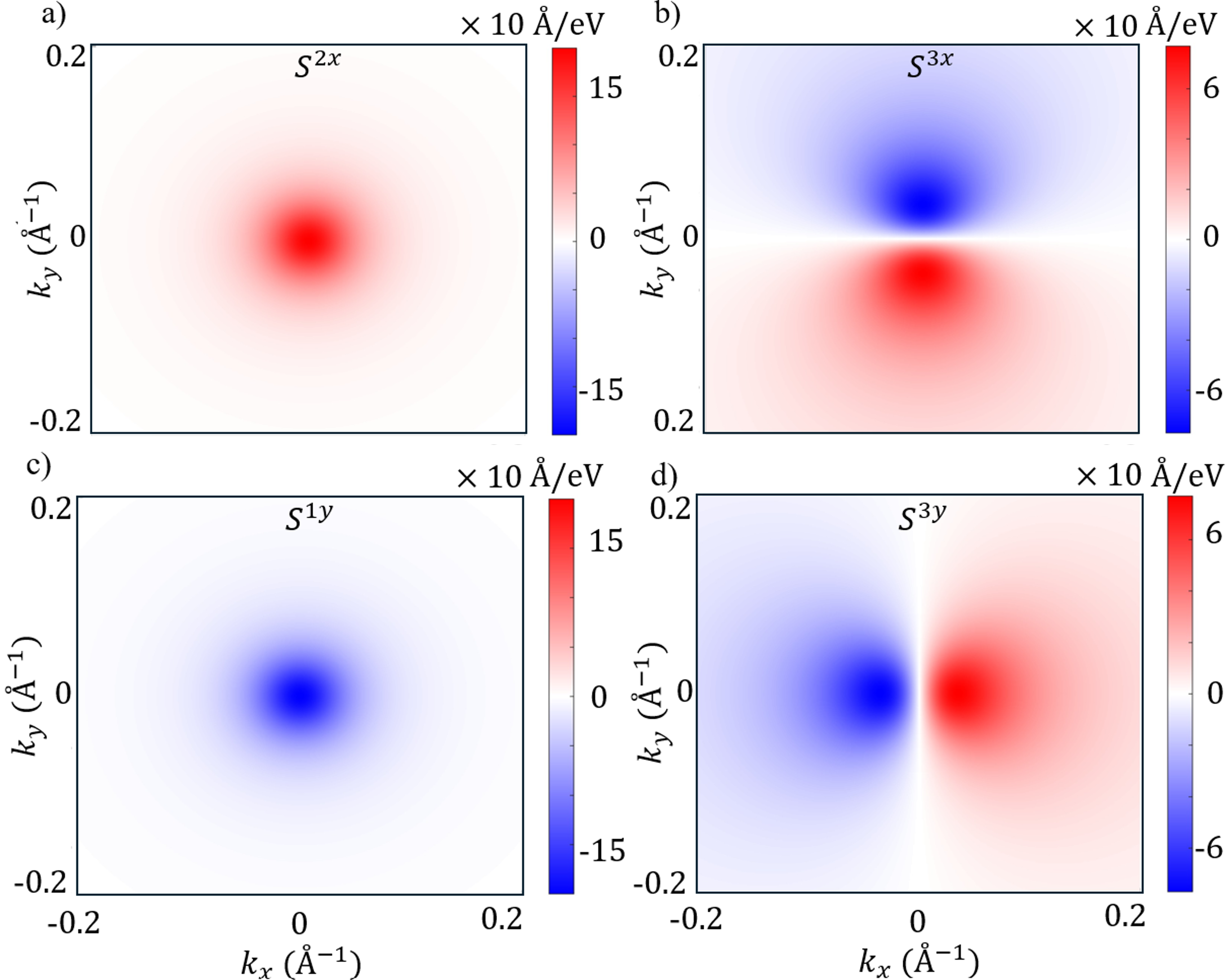}
    \caption{\textbf{Spin Dependent SBP distributions.} SBP $S^{\eta\alpha} (\mathbf{k})$ tensor distributions obtained from Eq.~(\ref{eq:deltaA}); here we have used $H_{\rm PT}^{(0)}$, see text. Note that $S^{1x}(\mathbf{k})$\ and $S^{2y} (\mathbf{k})$ vanishes for $H_{\rm PT}^{(0)}$. The in plane contributions are monopolar $S^{2x}(\mathbf{k})$ and $S^{1y}(\mathbf{k})$ [panel (a) and (c)] while the spins induced in $\hat{z}$ direction are dipolar $S^{3x} (\mathbf{k})$ and $S^{3y} (\mathbf{k})$ [panel (b) and (d)]}
    \label{Fig 2}
\end{figure}

At the core of the nonequilibrium collective quantum geometry is the SBP tensor. Unlike the BCP tensor $G^{\alpha \beta}_{n,\mathbf{k}}$ that comprises a product of two polar vectors $\mathbf{r} \otimes \mathbf{r}$, the SBP tensor $\mathcal{S}^{\eta\alpha}_{n,\mathbf{k}}$ is a product of a polar vector and an axial vector $\mathbf{r} \otimes \mathbf{s}$. As a result, $\mathcal{S}^{\eta\alpha}_{n,\mathbf{k}}$ depends on both the spin orientation and the spatial direction: e.g., $\mathcal{S}^{2x}(k_x,k_y)$ and $\mathcal{S}^{1y} (k_x,k_y)$ is monopolar (Fig.~\ref{Fig 2}a,c) while $\mathcal{S}^{3x}(k_x,k_y)$ and $\mathcal{S}^{3y}(k_x,k_y)$ is two-fold (Fig.~\ref{Fig 2}b,d).

Other quantum geometric quantities such as the momentum-space Berry curvature and quantum metric are similarly affected. Expanding to first order in $\mathbf{E}$ and $U_{\mathbf{k}}^\eta (\mathbf{E})$, we find the corrections to Berry curvature and quantum metric as $\tilde{\boldsymbol{\Omega}}_n(\mathbf{k}) = \boldsymbol{\Omega}_n^{(0)}(\mathbf{k}) + \delta \boldsymbol{\Omega}_n(\mathbf{k}, \mathbf{E})$ and $\tilde{\mathbf{g}}_n(\mathbf{k}) = \mathbf{g}^{(0)}_n(\mathbf{k}) + \delta \mathbf{g}_n(\mathbf{k}, \mathbf{E}) $ with: 
\begin{align}
    \delta {{\Omega}}^\gamma_n(\mathbf{k}, \mathbf{E}) &= \varepsilon^{\gamma\alpha\beta} \partial_\mathbf{k}^\alpha \left[
{ U_{\mathbf{k}}^\eta\mathcal{S}}_{n,\mathbf{k}}^{\eta\beta} \hspace{-0.5mm}+\hspace{-0.5mm}e {E}^\kappa {G}^{\kappa\beta }_{n,\mathbf{k}}\right],\label{fullQG}\\
       \delta{g}^{(\alpha\beta)}_n (\vec k, \mathbf{E})  &=\hspace{-3mm}\sum_{m,\epsilon_m\neq \epsilon_n} \hspace{-3mm}{\rm Re}[A^{\alpha}_{nm} ( U_{\mathbf{k}}^\eta\tilde{\mathcal{S}}^{\eta\beta }_{nm}+eE^\gamma \tilde{G}^{\gamma \beta}_{nm})], 
\end{align}
where $\varepsilon^{\gamma \alpha \beta}$ is the Levi-Civita tensor, the superscript braces $\mathcal{O}^{(\alpha\beta)}\equiv \mathcal{O}^{\alpha\beta}+\mathcal{O}^{\beta\alpha}$ denotes symmetric tensor component and we introduced a generalized interband SBP $\tilde{\mathcal{S}}^{\eta \beta}_{nm} = \sum_{c, \epsilon_c\neq\epsilon_m}s^\eta_{mc} A^\beta_{cn}/{\epsilon_{mc}} - \sum_{c, \epsilon_c\neq\epsilon_n}A^\beta_{mc} s^\eta_{cn} /{\epsilon_{cn}}$ and interband BCP $\tilde{G}^{\alpha\beta}_{nm} = \sum_{c, \epsilon_c\neq\epsilon_m}A^\alpha_{mc} A^\beta_{cn}/{\epsilon_{mc}} - \sum_{c, \epsilon_c\neq\epsilon_n} A^\beta_{mc} A^\alpha_{cn}/{\epsilon_{cn}}$ tensors. Here $\boldsymbol{\Omega}_n^{(0)}(\mathbf{k})$ and $\mathbf{g}^{(0)}_n(\mathbf{k})$ are the bare Berry curvature and quantum metric of the equilibrium bandstructure respectively. Both $ \delta \boldsymbol{{\Omega}}_n(\mathbf{k}, \mathbf{E})$ and $\delta{g}^{(\alpha\beta)}_n (\vec k, \mathbf{E})$ only occur at finite $\mathbf{E}$ underscoring its nonequilibrium nature.

It is interesting to note that the product  $r_{nm}^\alpha (\mathbf{k})s_{mn}^{\eta}(\mathbf{k})$ in Eq.~(\ref{eq:deltaA}) generates a Zeeman quantum geometry responsible for spin-related responses~\cite{Wang2025,Xiang2025} such as the Edelstein and spin Hall effects~\cite{Xiang2025}. Eq.~(\ref{fullQG}) demonstrates that for interactions in a nonequilibrium setting, the momentum-space and Zeeman quantum geometries become intertwined: as we will see, nonlinear charge responses can depend on both generators.

\vspace{2mm}
\textit{\color{blue} Exchange nonlinear Hall effect.} As an illustration of exchange driven nonequilibrium quantum geometry, we now examine the nonlinear Hall effect. We compute anomalous Hall currents as $\mathbf{j}_H =-(e^2/\hbar)\sum_{n\mathbf{k}} (\mathbf{E}\times\tilde{\boldsymbol{\Omega}}(\mathbf{k})) \, f_n (\mathbf{k})$. Substituting Eq.~(\ref{fullQG}) and isolating the exchange contributions, we obtain the exchange nonlinear   Hall (ENH) current as 
\begin{equation}
    {j}^\gamma_{\rm ENH}
\hspace{-0.5mm}=\hspace{-0.5mm} \frac{e^2}{\hbar} E^\alpha\hspace{-1mm}\sum_{n,\mathbf{k}}\left[(\partial^\gamma_\mathbf{k} f_n)U_{\mathbf{k}}^\eta(\mathbf{E})\mathcal{S}_{n,\mathbf{k}}^{\eta \alpha } \hspace{-0.5mm}-\hspace{-0.5mm}( \partial^\alpha_\mathbf{k} f_n)U_{\mathbf{k}}^\eta(\mathbf{E})\mathcal{S}_{n,\mathbf{k}}^{\eta \gamma}\right], 
\label{ENH}
\end{equation}
where we have noted that $\mathcal{S}_{n,\mathbf{k}}^{\beta \eta} U_{\mathbf{k}}^\eta (\mathbf{E})$ is periodic in the Brilluoin zone leading to a vanishing boundary term. As a result, ENH is a Fermi surface effect in the same fashion as other nonlinear Hall effects such as the intrinsic nonlinear Hall (INH) effect~\cite{Gao2014,Liu2021,Wang2021} and Berry curvature dipole (BCD) \cite{Sodemann2015}. Notice that $U_{\mathbf{k}}^\eta (\mathbf{E})$ is directly proportional to $\mathbf{E}$ making ${j}^\gamma_{\rm ENH}$ a second-order nonlinearity.

\begin{figure}
\centering
\includegraphics[width=0.5\textwidth]{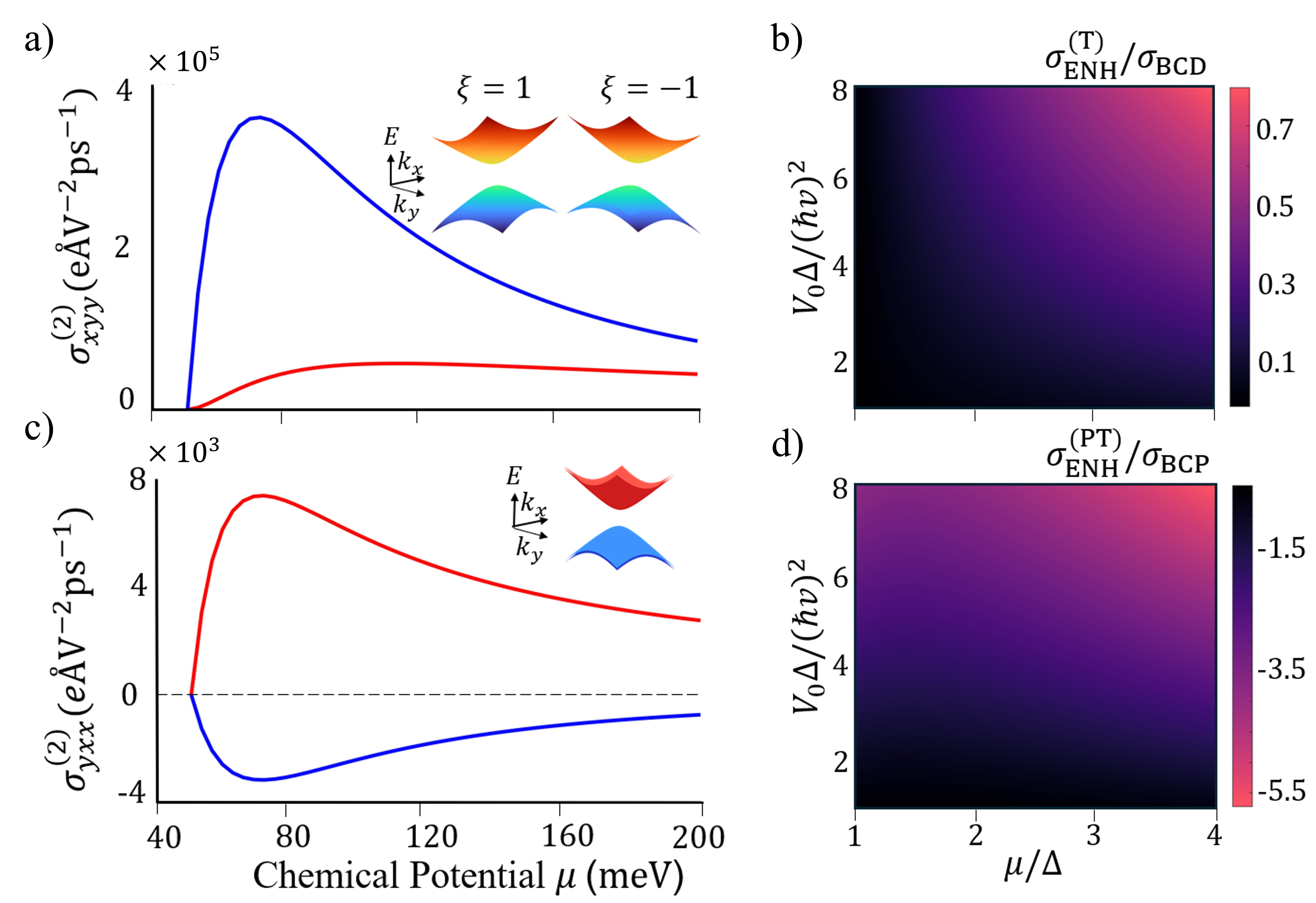}
\caption{\textbf{ENH in noncentrosymmetric materials.} (a) ENH second-order conductivity $\sigma_{\rm ENH}^{(\rm T)}$ (red) and BCD $\sigma_{\rm BCD}$ (blue) for a non-magnetic noncentrosymmetric $H_{\rm T}^{(0)} (\mathbf{k})$ (see text). (b) The ratio of $\sigma_{\rm ENH}^{(\rm T)}$ to  $\sigma_{\rm BCD}$ grows with interaction strength and chemical potential becoming competitive at larger chemical potentials and interaction strength. The relaxation time scale for plots (a) and (b) are chosen to be $\tau=0.5\ \text{ps}$. (c) Contrasting $\sigma_{\rm ENH}^{(\rm PT)}$ (red) and $\sigma_{\rm INH}$ (blue) for a PT-symmetric magnetic $H_{\rm PT}^{(0)}(\mathbf{k})$ (see text). (d) Ratio of $\sigma_{\rm ENH}^{(\rm PT)}$ against $\sigma_{\rm INH}$ indicates ENH is competitive across a wide parameter space. Other model parameters are $\Delta = 50\, {\rm meV}$, $v_{x,y} = 4 \times 10^5 \, {\rm m} {\rm s}^{-1}$, $\alpha/\hbar v = 0.1$, $q_s=0.2\ \text{nm}^{-1}$ for both $H_{\rm T}^{(0)}$ and $H_{\rm PT}^{(0)}$ in panel (a) and (c) respectively.}
\label{Fig 3}
\end{figure}

Writing ${j}^\gamma_{\rm ENH} = \sigma_{\rm ENH}^{\gamma \alpha \beta} E^\alpha E^\beta$, we separate out ENH second-order nonlinear conductivity into two distinct contributions $\sigma_{\rm ENH} = \sigma^{\rm{(T)}}_{\rm ENH}+ \sigma^{\rm{(PT)}}_{\rm ENH}$. Here ${\sigma}^{\rm{(PT)}}_{\rm ENH}$ is PT-even (but T-odd) reading as    
\begin{equation}
   [{\sigma}^{({\rm PT})}_{\rm ENH}]^{\gamma\alpha\beta}\hspace{-0.5mm}=\hspace{-0.5mm} \frac{e^3}{\hbar}\hspace{-1mm}\sum_{nn'\atop\mathbf{kk'}}\hspace{-0.5mm}(\partial^\gamma_\mathbf{k} f^{\mathbf{k}}_n) \mathcal{S}^{\eta\alpha}_{n,\mathbf{k}} V_{\mathbf{k'-k}}\mathcal{S}^{\eta\beta}_{n',\mathbf{k'}}{f_{n'}^{\mathbf{k'}}} - \gamma \leftrightarrow \alpha,
\label{sigmaSBPPT}
\end{equation}
Notice that Eq.~(\ref{sigmaSBPPT}) does not scale with $\tau$ and as such mirrors that of the INH. In contrast, $\sigma^{\rm{(T)}}_{\rm ENH}$ is T-even (but PT-odd) reading as
\begin{equation}
[{\sigma}^{({\rm T})}_{\rm ENH}]^{\gamma\alpha\beta}\hspace{-1mm}=\hspace{-1mm} \frac{e^3\tau}{\hbar^2}\hspace{-1mm}\sum_{nn'\atop\mathbf{kk'}}\hspace{-0.5mm}\mathcal{S}^{\eta\alpha}_{n,\mathbf{k}}V_{\mathbf{k'-k}}s^{\eta \mathbf{k}'}_{n'n'}\partial^\gamma_\mathbf{k} f^{\mathbf{k}}_n\partial^\beta_\mathbf{k'}f^{\mathbf{k'}}_{n'} - \gamma \leftrightarrow \alpha,
\label{sigmaSBPT}
\end{equation}
which scales linearly with $\tau$; this mirrors the scaling of BCD nonlinear hall effect at low frequencies. Notice that $\sigma_{\rm ENH}$ is always transverse to applied electric fields~\cite{Souza2022}. As a result, much like other nonlinear Hall effects, this imposes strong point-group symmetry constraints, e.g., in two-dimensional materials, ENH requires broken rotational symmetry~\cite{Sodemann2015}. 

Several comments are in order. First, we note that the SBP for $\eta =0$ (i.e. $\mathcal{S}^{\alpha0}$) identically vanishes: ENH directly depends on spin density accumulation. As a result, ENH in Eq.~(\ref{sigmaSBPPT}) and (\ref{sigmaSBPT}) is suppressed for materials without spin-orbit coupling or where spin is a good quantum number. This makes strong-orbit coupling materials where inversion breaking is built-in to spin order ideal candidates for ENH.

Second, we note that ENH in Eq.~(\ref{sigmaSBPPT}) and (\ref{sigmaSBPT}) demonstrates that nonlinear Hall effects are not purely tied to the geometry of the equilibrium bandstructure: SBP provides exchange corrections that are only activated out-of-equilibrium. This is particularly striking for the nonlinear Hall effects in T-symmetric materials that scale as $\tau$ and have been thought to be dominated by the BCD~\cite{Sodemann2015}: $\mathcal{D}^{\alpha\beta}_{\rm bcd} = \sum_{n,\mathbf{k}}\partial^\alpha_\mathbf{k} \Omega^\beta_n(\mathbf{k}) f_n(\mathbf{k})$; $\mathcal{D}^{\alpha\beta}_{\rm bcd}$ is a ground state property. In contrast, we find $[{\sigma}^{({\rm T})}_{\rm ENH}]$ in Eq.~(\ref{sigmaSBPT}) scales linearly with $\tau$. Nevertheless, it depends on SBP, interaction strength, and the nonequilibrium nature of the many-body state. This complicates the use of $\tau$ dependence as a means of extracting groundstate BCD~\cite{ma2019,Kang2019}.

\vspace{2mm}
\textit{\color{blue} ENH in noncentrosymmetric quantum materials}. To illustrate the ENH, we examine a minimal spin model where inversion symmetry is broken. We first focus on the T-symmetric case with an effective mean-field Rashba Hamiltonian~\cite{Sodemann2015} : $H^{(0)}_{\rm T}(\mathbf{k}) = \hbar v_x k_x s_y - \xi \hbar v_y k_y s_x+ \Delta s_z + \xi \alpha k_y$ where $\xi = \pm1$ describes two inequivalent valleys (see inset Fig \ref{Fig 3}a), $v_{x,y}$ are velocities, $\Delta$ is a gap. $H^{(0)}_{\rm T}(\mathbf{k})$ has been widely used to describe BCD~\cite{Sodemann2015} and mirrors the behavior of Dirac surface states in topological crystalline insulators e.g., SnTe \cite{Hsieh2013}. The tilt parameter $\alpha$ breaks rotational symmetry activating a nonlinear Hall effect described by the element $\sigma_{xyy}^{(2)} = -\sigma_{yxy}^{(2)}$. In what follows, we compute $\sigma_{xyy}^{(2)}$ to lowest order in $\alpha$ at zero temperature 

Adopting a simple contact interaction $V_{\mathbf{k-\mathbf{k}'}} = V_0$, we plot the T-even ENH second order conductivity in Eq.~(\ref{sigmaSBPT}) (red) and the BCD nonlinear Hall conductivity (blue)~\cite{Sodemann2015} in Figure \ref{Fig 3}a where we used ${\sigma}^{\gamma \alpha\beta}_{\rm BCD} =(e^3 \tau/\hbar^2 )\epsilon^{\gamma\beta\rho}\mathcal{D}^{\alpha\rho}_{\rm bcd}$~\cite{Sodemann2015}. Here we have estimated $V_0 = e^2/(4\pi\epsilon_0 q_s)$ where we used the screening parameter $q_s =0.2 \, {\rm nm}^{-1}$ as an illustration; $\epsilon_0$ is the vacuum permittivity. Fig.~\ref{Fig 3}a shows that ENH becomes competitive at large $\mu$, while BCD dominates at low chemical potential $\mu$. In general, we find $\sigma^{(2)}_{\rm ENH}/\sigma^{(2)}_{\rm BCD}$ grows at large chemical potential and interaction strength but becomes suppressed at low chemical potential and interaction strength as shown in Fig \ref{Fig 3}b. This behavior underscores the competitive role ENH plays in accounting for nonlinear Hall conductivities.

A similar behavior arises in PT symmetric materials. To see this we examine a minimal spin model~\cite{Zhang2016}: $H^{(0)}_{\rm PT}(\mathbf{k}) = \tau_y(\hbar v_x k_x s_x + \hbar v_y k_y s_y+ \Delta s_z) + \alpha k_y$ where $\tau_y$ describes an orbital degrees of freedom. PT symmetry ensures Kramers degeneracy at each $\mathbf{k}$~\cite{Zhang2016} zeroing spin density in equilibrium. Nevertheless, spin is not a good quantum number in $H^{(0)}_{\rm PT}(\mathbf{k})$, allowing applied electric fields to induce nonequilibrium spin accumulation~\cite{Xiang2025} and ENH.

In PT symmetric materials, both the BCD nonlinear Hall effect as well as $\sigma^{(\rm T)}_{\rm ENH}$ vanish. Instead, here we find $\sigma^{(\rm PT)}_{\rm ENH}$ and the intrinsic nonlinear Hall effect~\cite{Gao2014,Liu2021,Wang2021} manifest; here we computed INH directly as ${\sigma}_{\rm INH}^{\gamma\alpha\beta}= (e^3/\hbar)\sum_{n,\mathbf{k}}  G^{\beta\alpha}_{n,\mathbf{k}}\partial^\gamma_\mathbf{k} f^\mathbf{k}_n-(\gamma\leftrightarrow\alpha)$~\cite{Gao2014,Liu2021,Wang2021}. We plot $\sigma^{(\rm PT)}_{\rm ENH}$ and the INH second order conductivity in Fig.~\ref{Fig 3}c and find that $\sigma^{(\rm PT)}_{\rm ENH}$ can provide large contributions that can dominate over INH; it also has an opposite sign. We note while both generally exhibit a similar nonmonotoic behavior away from the band edge, at large $\mu$ they scale differently: $\sigma^{(\rm PT)}_{\rm ENH} \sim 1/\mu$ while ${\sigma}_{\rm INH}\sim 1/\mu^2$. This difference in $\mu$ dependence enables to differentiate between the mechanisms. We note that the precise dominance of INH vs ENH can depend sensitively on parameter values and the location of the chemical potential, Fig.~\ref{Fig 3}d.

In the nonlinear Hall effect, electron-electron interactions are often thought to play a {\it passive} role entering as part of a frozen background meanfield. Our work demonstrates that exchange interactions can instead play a {\it dynamical} role generating ENH from nonequilibrium exchange interactions. ENH is not locked to the momentum-space quantum geometry of the equilibrium ground state but instead to how the many-body state is pushed out-of-equilibrium (e.g., by an electric field). While here we have focused on the second-order charge nonlinear Hall effect, a range of other responses may also be similarly affected by such nonequilibrium exchange interactions. For instance, we anticipate that nonequilibrium exchange may provide $\mathbf{E}$ induced changes to the group velocity impacting for e.g., longitudinal nonlinear transport and extrinsic disorder mediated mechanisms such as skew-scattering~\cite{konig2019,Du2019,Isobe2020,Ma2023}. Perhaps most exciting are nonlinear photocurrents that are often described in terms of momentum-space interband quantum geometry of Bloch states~\cite{Ahn2022,MaQ2023}. Similar nonequilibrium exchange interactions may also be operative for nonlinear photocurrents tracking a nonequilibrium many-body quantum geometry.

{\it \color{blue} Equal Contribution:} $\dagger$ indicates that John Tan and Oles Matsyshyn contributed equally to this work.

\textit{\color{blue} Acknowledgements.} We are grateful for useful conversations with Mark Rudner and Xu Yang. J.C.W.S was supported by the Singapore Ministry of Education (MOE) AcRF Tier 2 grant MOE-T2EP50222-0011 and Tier 3 grant MOE-MOET32023-0003 “Quantum Geometric Advantage”. G. V. was supported by the Ministry of Education, Singapore, under its Research Centre of Excellence award to the Institute for Functional Intelligent Materials (I-FIM, Project No. EDUNC-33-18-279-V12).

\end{document}